\documentclass[useAMS,usenatbib]{mn2e}

\usepackage{newtxtext,newtxmath}

\usepackage{graphicx}
\usepackage{float}
\usepackage{color}
\usepackage{ae,aecompl}
\usepackage{amsmath}
\usepackage{hyperref}

\newcommand{\aj}{AJ}			
\newcommand{\apj}{ApJ}			
\newcommand{\apjl}{ApJL}		
\newcommand{\apjs}{ApJS}		
\newcommand{\aap}{A\&A}			
\newcommand{\aaps}{A\&AS}		
\newcommand{\mnras}{MNRAS}		
\newcommand{\pasp}{PASP}		
\newcommand{\na}{New Astronomy} 
\newcommand{\cjaa}{Chinese Journal of Astronomy \& Astrophysics}
\newcommand{\pasa}{Publications of the Astronomical Society of Australia}

\newcommand{\Ha}{H$\alpha$}

\newcommand{\Oii}{[O~{\sc ii}] $\lambda$3727}

\usepackage[usenames,dvipsnames]{xcolor} 

\def\aap{Astronomy \& Astrophysics}
\def\apj{The Astrophysical Journal} 
\def\mnras{Monthly Notices of the Royal Astronomical Society}

\title[The S-PLUS: a star/galaxy classification based on a Machine Learning approach]{The S-PLUS: a star/galaxy classification based on a Machine Learning approach}

\author[]{M. V. Costa-Duarte$^{1}$\thanks{E-mail: mvcduarte@gmail.com}, L. Sampedro$^{1}$, A. Molino$^{1}$, H. S. Xavier$^{1}$, 
\newauthor{F. R. Herpich$^{1}$,
A. L. Chies-Santos$^{2}$, C. E. Barbosa$^{1,3}$, A. Cortesi$^{1,4}$, W. Schoenell$^{2}$,}
\newauthor{A. Kanaan$^{5}$, T. Ribeiro$^{6}$, C. Mendes de Oliveira$^{1}$, S. Akras$^{4,7}$, A. Alvarez-Candal$^{7}$,}
\newauthor{C. L. Barbosa$^{8}$, J. L. N. Castell\'on$^{9,10}$, P., Coelho$^{1}$, M. L. L. Dantas$^{1}$, R. Dupke$^{7}$,}
\newauthor{ A. Ederoclite$^{1}$, A. Galarza$^{7}$, T. S. Gon\c calves$^{4}$, J. A. Hernandez-Jimenez$^{1,11}$,} 
\newauthor{Y. Jim\'enez-Teja$^{7}$, A. Lopes$^{7}$, P. A. A. Lopes$^{4}$, R. Lopes de Oliveira$^{12,7}$,}
\newauthor{J. L. Melo de Azevedo$^{13}$, L. M. Nakazono$^{1}$, H. D. Perottoni$^{1}$, C. Queiroz$^{14}$, K. Saha$^{15}$,} 
\newauthor{L. Sodr\'e Jr.$^{1}$, E. Telles$^{7}$, R. C. Thom de Souza$^{16}$}\\
$^{1}$Instituto de Astronomia, Geof\'isica e Ci\^encias Atmosf\'ericas, Universidade de S\~ao Paulo, 05508-090, S\~ao Paulo, Brazil\\
$^{2}$Departamento de Astronomia, Instituto de F\'isica, Universidade Federal do Rio Grande do Sul, Porto Alegre, Brazil\\
$^{3}$Steward Observatory, University of Arizona, 933 N Cherry Ave, Tucson, AZ 85719, United States\\
$^{4}$Valongo Observatory, Federal University of Rio de Janeiro, Ladeira Pedro Antonio 43, Saude Rio de Janeiro, RJ, 20080-090, Brazil \\
$^{5}$ Departamento de F\'isica, Universidade Federal de Santa Catarina, Florian\'{o}polis, SC, 88040-900, Brazil \\
$^{6}$ NOAO, P.O. Box 26732, Tucson, AZ 85726\\
$^{7}$ Observat\'orio Nacional / MCTIC, Rua General Jos\'e Cristino 77, Rio de Janeiro, RJ, 20921-400, Brazil\\
$^{8}$Dept. of Physics, Centro Universit\'ario da FEI Av. Humberto de Alencar Castelo Branco, 3972. 09850-901\\ S\~ao Bernardo do Campo - SP, Brazil.\\
$^{9}$Instituto de Investigacion Multidisciplinario en Ciencia y Tecnolog\'ia, Universidad de La Serena. Benavente 980, La Serena, Chile\\
$^{10}$Departamento de F\'isica y Astronom\'ia, Universidad de La Serena. Avenida Juan Cisternas 1200, La Serena, Chile.\\
$^{11}$Departamento de Ciencias F\'isicas, Universidad Andr\'es Bello, Fern\'andez Concha 700, Las Condes, Santiago, Chile\\
$^{12}$Departamento de F\'isica, Universidade Federal de Sergipe, Av. Marechal Rondon, S/N, 49000-000, S\~ao Crist\'ov\~ao, SE, Brazil\\
$^{13}$Intituto de Ci\^encias Matem\'aticas e de Computa\c c\~ao - Universidade de S\~ao Paulo (ICMC-USP), S\~ao Paulo, Brazil\\
$^{14}$Departamento de F\'isica Matem\'atica, Instituto de F\'isica, Universidade de S\~{a}o Paulo, SP, Rua do Mat\~{a}o 1371, S\~{a}o Paulo, Brazil\\
$^{15}$Inter-University Centre for Astronomy and Astrophysics, PostBag 4, Ganeshkhind, Pune-411007, India\\
$^{16}$Universidade Federal do Paran\'a, Campus Jandaia do Sul, Rua Dr. Jo\~ao Maximiano, 426, Jandaia do Sul-PR, 86900-000, Brazil.}
\begin{document}
\date{Accepted . Received ; in original form }

\pagerange{\pageref{firstpage}--\pageref{lastpage}} \pubyear{2018}

\maketitle

\label{firstpage}

\begin{abstract}
We present a star/galaxy classification for the Southern Photometric Local Universe Survey (S-PLUS), based on a Machine Learning approach: the Random Forest algorithm. We train the algorithm using the S-PLUS optical photometry up to $r$=21, matched to SDSS/DR13, and morphological parameters. The metric of importance is defined as the relative decrease of the initial accuracy when all correlations related to a certain feature is vanished. In general, the broad photometric bands presented higher importance when compared to narrow ones. The influence of the morphological parameters has been evaluated training the RF with and without the inclusion of morphological parameters, presenting accuracy values of 95.0\% and 88.1\%, respectively. Particularly, the morphological parameter {\rm FWHM/PSF} performed the highest importance over all features to distinguish between stars and galaxies, indicating that it is crucial to classify objects into stars and galaxies. We investigate the misclassification of stars and galaxies in the broad-band colour-colour diagram $(g-r)$ versus $(r-i)$. The morphology can notably improve the classification of objects at regions in the diagram where the misclassification was relatively high. Consequently, it provides cleaner samples for statistical studies. The expected contamination rate of red galaxies as a function of the redshift is estimated, providing corrections for red galaxy samples. The classification of QSOs as extragalactic objects is slightly better using photometric-only case. An extragalactic point-source catalogue is provided using the classification without any morphology feature (only the SED information) with additional constraints on photometric redshifts and {\rm FWHM/PSF} values.  
\end{abstract}

\begin{keywords}
catalogues - galaxies: photometry - surveys: multiwavelength
\end{keywords}

\section{Introduction}
\label{intro}


Several astronomical surveys need to face the challenge of classifying observed sources in terms of their astronomical nature, i.e., stars, galaxies, QSOs, planetary nebulae, supernovae, among others \citep[][and references therein]{Robertsonetal2015}. The classification of objects in survey catalogues could speed up considerably studies in several fields, as evidenced by the widely successful Sloan Digital Sky Survey \citep[SDSS,][]{Yorketal2000, Abolfathietal2017} and the myriad of queries and samples made out of it. Studies on classification of astronomical sources are needed to estimate, for instance, contamination rates even for surveys that are conceived to be focused on a particular object/science case. This classification becomes more crucial for photometric surveys, where the spectral information is not available to precisely classify objects.

Apart from deep pencil-like surveys which cover small areas in the sky and large extragalactic volumes, the current generation of extragalactic surveys generally covers large areas in the sky and low/intermediate redshift ranges, e.g., Javalambre Physics of the Accelerating Universe Astrophysical Survey \citep[J-PAS,][]{Benitezetal2014}, Dark Energy Survey \citep{DES}, Euclid \citep{EUCLID} and Large Synoptic Survey Telescope \citep[LSST, ][]{LSST}, S-PLUS \citep{MendesdeOliveiraetal2019} and J-PLUS \citep{Cenarroetal2018}. Other surveys focused on Galactic and stellar science also have the necessity of classifying objects to obtain pure stellar samples, such as Pan-STARSS \citep{PANSTARSS} and Gaia survey \citep{GAIA}.

This task demands a particular approach since some of the aforementioned surveys have relatively large number of photometric bands and consequently it becomes a high-dimensional problem. Several classification algorithms have included photometric colours \citep[e.g.,][]{Polloetal2010, KovacsSzapudi2016} and/or morphological parameters, such as extension and concentration index calculated, for instance, by \texttt{SEXTRACTOR} \citep{SEXTRACTOR}. Other works implemented a more complex approach on astronomical images, such as moment indices and granulometry \citep[][]{Candeasetal1997, Mooreetal2006}. As demonstrated in \cite{Pimbbletetal2001}, among several parameters, both the Full Width at Half Maximum ({\rm FWHM}) and \texttt{SEXTRACTOR}'s stellarity parameter, provide a reliable morphological proxy to separate stars and galaxies.

Recently, several regressions and classification problems in Astronomy have started to be approached using an assortment of Machine Learning (hereafter ML) algorithms. They represent one of the main branches of this new generation of tools. They are able to identify complex nonlinear behavior in the multi-dimensional feature space, providing accurate solutions for several questions in Astronomy. Photometric redshifts \citep{Cavuotietal2014, Sadehetal2016, Gomesetal2017, Bilickietal2017} and galaxy morphological classification \citep{SchutterShamir2015, Barchietal2017, Hockingetal2017, DominguezSanchezetal2018} are some examples which use ML. Particularly, the star/galaxy classification has been approached with several ML techniques, such as Support Vector Machines (SVMs) \citep{Solarzetal2012, Fadelyetal2012}, Decision Trees and Random Forest (hereafter RF) \citep{Breiman2001, Odewahnetal2004, Balletal2006, Vasconcellosetal2011}, Artificial Neural Networks (ANNs) \citep[][]{Qinetal2003, Boraetal2009, KimBrunner2016}. In particular, the \texttt{SExtractor} software is based on ANNs and widely used in astronomical images to classify stars and galaxies. Moreover, there are other techniques that tackle the star/galaxy separation in the literature, focused on SED-fitting \citep{Wolfetal2004, Robinetal2007, Preethietal2014}, Bayesian statistics \citep[][]{Herionetal2011, Molinoetal2014, LopezSanJuanetal2018} and Principal Component Analysis (PCA) \citep[][]{Cabanacetal2002, Soumagnacetal2015}. All these techniques have their strength and limitation, depending on the problem approached. Recently, \cite{Machadoetal2016} evaluated a set of ML algorithms for the star/galaxy classification problem. The authors demonstrated that ANNs and RF performed better when compared to other methods, such as k-Nearest Neighbour \citep{BathiaVandana2010} and Naive Bayes \citep{Jiangetal2007}. In particular, the RF algorithm has some features which are suitable for the star/galaxy classification problem, i.e., the overfitting problem can be reduced and errors and probabilities are reliably calculated by averaging over a large enough number of trees \citep{RobnikSikonja2004, Boineeetal2005}.

Here we present a star/galaxy separation technique based on the RF algorithm and its application to the S-PLUS data. This paper is organized as follows. In Section $\ref{data}$, we briefly introduce the S-PLUS survey, for which the work presented in this paper is meant to be applied to, given its filter system and photometric-depth. In Section $\ref{methods}$, we present the RF algorithm which has been used to classify objects using the multi-band photometry and morphology as input features. Section $\ref{results}$ shows the performance (completeness, purity and misclassification) of the RF algorithm as a function of the apparent r-band magnitude and broad band colours. Section \ref{sec.application} presents some applications of the developed star/galaxy separation algorithm: a compiled point-source extragalactic catalogue. Then, in Section $\ref{conclusions}$, our conclusions and discussions on the ML approach and results are presented.

\vspace{0.2cm}

The S-PLUS photometry presented here are in AB system. We have adopted the following cosmological parameters: $H_{0}$ = 70 km$s^{-1}$ Mpc$^{-1}$, $\Omega_{M}$ = 0.3, $\Omega_{\Lambda}$ = 0.7 and $\Omega_{K}$=0.0.

\section{Database}
\label{data}

\subsection{The S-PLUS survey}

The Southern Photometric Local Universe Survey \footnote{\url{http://www.splus.iag.usp.br}} \citep[S-PLUS;][]{MendesdeOliveiraetal2019} is an ongoing photometric survey focused on Galactic science cases and will observe 9,000 sq. deg. of the Southern Hemisphere. The survey employs a dedicated 0.8m fully robotic telescope (T80-South) at Cerro Tololo, Chile. S-PLUS uses a 5 broad- and 7 narrow bands filter system in the optical, strategically designed to cover some stellar and galactic spectral features in the local Universe, such as metal-dependent features (e.g. Ca H+K and G band) and emission lines (e.g. \Ha\ and \Oii). Table \ref{tab.filter_system} indicates their central wavelengths and widths. 

The S-PLUS Data Release 1 (S-PLUS/DR1)\footnote{\url{https://datalab.noao.edu/splus/index.php}} has observed the SDSS Stripe 82 (S82) region which covers the region at the Celestial Equator between $-1.25^{o}$$<$$\delta$$<$$+1.25^{o}$ and $-60^{o}$$<$$\alpha$$<$$+60^{o}$ \citep{Jiangetal2014}. The S-PLUS/DR1 data is composed by 170 fields, each one covering 1.96 sq. deg., and more than 2 million of objects up to $r$=21 mag.  

In order to characterize the apparent morphology of objects, we used the following morphological parameters: the ellipticity defined as $b$/$a$ (being $a$ and $b$ are the semi-major and semi-minor axes, respectively), a concentration proxy C = $r_{AUTO}$ - $r_{PETRO}$, being the $r$ magnitude using the AUTO (from SExtractor) and Petrosian \citep{Petrosian1976} formalism, and the full width at half maximum divided by the mean {\rm PSF} (Point Spread Function) of the field. Our representative {\rm PSF} definition consists of {\rm FWHM} 5-th percentile value of objects in the magnitude range 14$<$$r$$<$17, taking the bright magnitude range dominated by stars and relatively small {\rm FWHM} in the field. All these parameters have been calculated from the detection images (see Sampedro et al in prep.). Since the {\rm FWHM} is an observation-dependent parameter and so it is expected to vary from one field to another, so we decide to normalize this quantity by the {\rm PSF} of the point-like source in the field. This approach allows us to define a more stable estimate of the {\rm FWHM} of sources through the entire survey. In this way, point-like sources present values close to unity while extended sources have larger values.

Our initial magnitude-limited sample was extracted from the S-PLUS database between 14$<$$r$$<$21 and matched with the SDSS Stripe 82 database, resulting in $\sim$924k objects in common for both catalogues. This photometric limit is defined to guarantee the balance between the photometric quality and depth in all bands. For our sample characterization, we selected all 12 S-PLUS bands (\texttt{AUTO} magnitudes) and the morphological parameters ({\rm FWHM/PSF}, $b$/$a$ and concentration) as input features for the ML algorithm. All objects in our initial sample have the SDSS/DR13 star/galaxy classification \citep{Albaretietal2017}. Furthermore, the SDSS Stripe 82 observations are deeper than S-PLUS data (SDSS data is complete down to $r$=24.6), so we have decided to rely on the SDSS star/galaxy classification as true for objects in this region and used its classification to train our algorithm. Thus, our goal in this work is to validate our classification by reproducing the SDSS classification from Stripe 82 for the S-PLUS survey and apply it on other S-PLUS fields in the future.
\begin{table}
\caption{{\small Central wavelengths and widths of the S-PLUS filter system. For further details, we refer the reader to Mendes de Oliveira et al. (2019)}}
\begin{center}
\label{tablefilter}
\begin{tabular}{lccc}
\hline
\hline
Filter	&	$\lambda_{\rm eff} $	&	$\Delta \lambda$ \\	
Name	&	[\AA]	&	[\AA] & Spectral feature \\	\hline
uJAVA & 3574 & 330 & Javalambre $u$ \\ 
F0378 & 3771 & 151 & $[\mathrm{O}\,\textsc{ii}]$ \\ 
F0395 & 3941 & 103 & Ca H+K \\ 
F0410 & 4094 & 201 & H$\delta$ \\ 
F0430 & 4292 & 200 & G-band \\ 
gSDSS & 4756 & 1536 & SDSS-like $g$ \\ 
F0515 & 5133 & 207 & Mgb Triplet \\ 
rSDSS & 6260 & 1462 & SDSS-like $r$ \\ 
F0660 & 6614 & 147 & H$\alpha$ \\ 
iSDSS & 7692 & 1504 & SDSS-like $i$ \\ 
F0861 & 8611 & 408 & Ca Triplet \\ 
zSDSS & 8783 & 1072 & SDSS-like $z$ \\ 
\hline
\hline
\end{tabular}
\end{center}
\label{tab.filter_system}
\end{table}   


\subsection{Training and Test samples}

\begin{figure}
\includegraphics[scale=0.3]{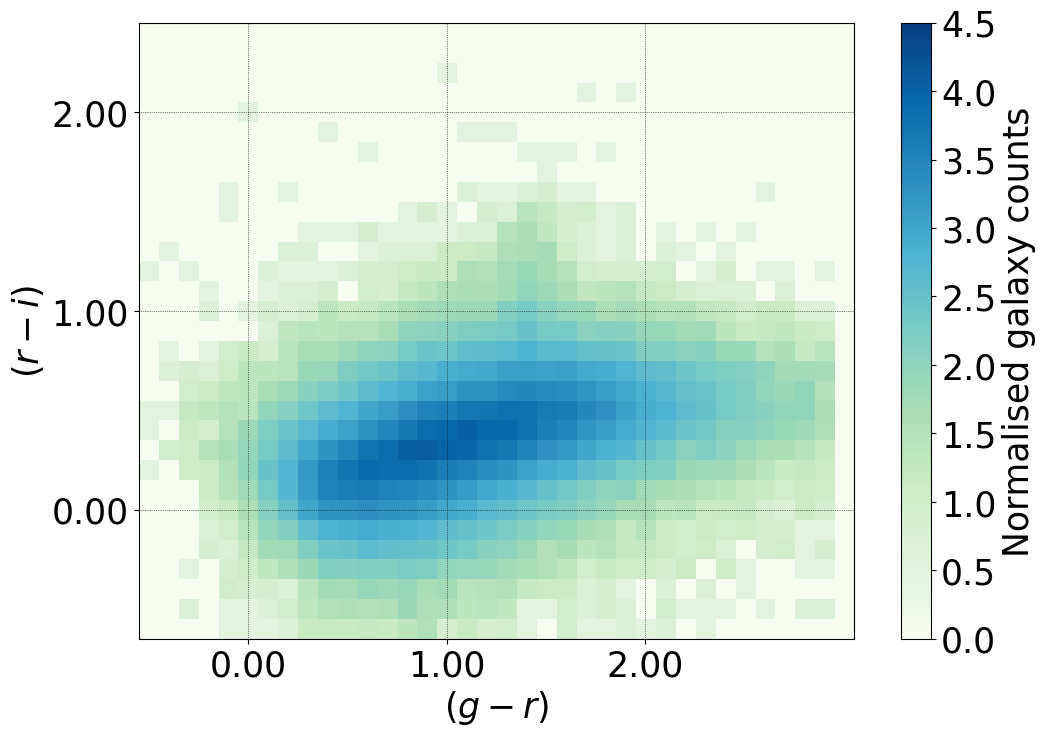}
\includegraphics[scale=0.3]{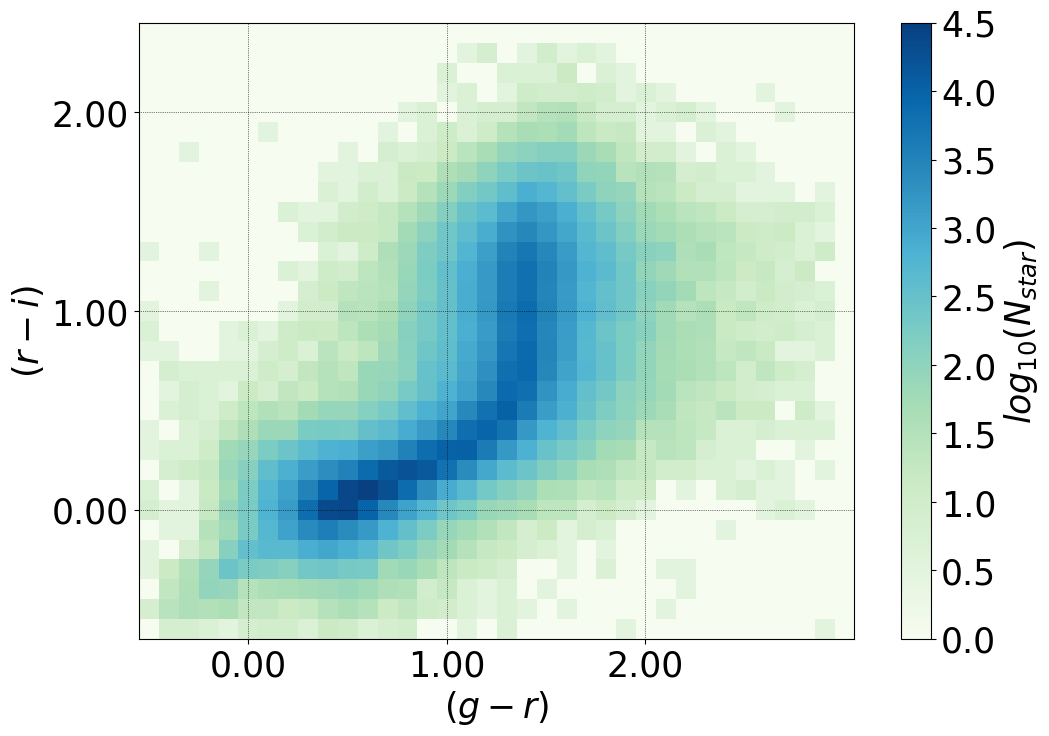}
\caption{The \textit{loci} of galaxies (upper panel) and stars (lower panel) in the colour-colour space $(g-r)$ versus $(r-i)$ of our initial sample. The classification shown here was extracted from the SDSS database.}
\label{fig.ngal_nstar}
\end{figure}

As a rule of thumb, supervised ML Algorithms are evaluated using an independent but representative sample compared to the training sample, called the test sample. The initial sample has been randomly split into training and test samples comprised of 724k and 200k objects respectively, to evaluate/validate the algorithm performance. Figure \ref{fig.train_test_samples} shows the apparent magnitude ($r$ band) distribution of initial sample (upper panel). It shows the dominance of stars/galaxies at the bright/faint edge of our sample. The lower panel shows the fraction of objects as a function of missing bands, being 59\% of the objects have all S-PLUS magnitudes measured, 20\% of objects present one missing band and 21\% have more than one missing band. We also characterize the distributions of stars and galaxies in the colour-colour space of the initial sample together. Figure \ref{fig.ngal_nstar} shows the \textit{loci} of stars and galaxies in the colour-colour space using broad bands, adopting $(g-r)$ and $(r-i)$. Stars and galaxies present distinct distributions, however, there is an overlapping \textit{locus}. This region may decrease our classification performance since there are objects from both classes occupying the same \textit{locus}. On the other hand, it is important to mention that it shows simplified 2-dimensional distributions of stars and galaxies in colour-colour space. The RF analysis works in a higher-dimensional space, including more features than only two photometric colours.


\section{Methodology}
\label{methods}

In this section, we describe the RF algorithm and how it is employed to classify objects as stars and galaxies based on S-PLUS magnitudes and morphological features.

\subsection{Random Forest Algorithm}
\label{sec.RF}

A diversity of topics can be investigated in Astronomy using ML such as galaxy morphology in seeing-limited images \citep{HuertasCompanyetal2009}, classification of stars, galaxies and active nuclei \citep{ZhangZhao2005} and photometric redshift estimates \citep{Wadadekar2005}. Particularly, the RF is frequently used in the astronomical context, dealing with high-dimensional data to classify or make regressions \citep[e.g. ][]{Gaoetal2009, OKeefeetal2009, Vasconcellosetal2011, Machadoetal2016, BaronPoznanski2017}. The RF package from the library \texttt{scikit-learn} \footnote{\url{http://scikit-learn.org/}} in Python \citep{SKLEARN} was chosen for this work.

Here we briefly describe the RF algorithm, in particular, how the trees are built and trained \citep[for more details, see][]{Breiman2001, RaileanuStoffel2004}. In the training procedure, the RF algorithm uses the training sample $\mathcal{X}$ with $M$ objects to create a number of trees (\texttt{$n_{\rm tree}$}) bootstrapped subsamples $\mathcal{S}$ with the same size as the initial sample $\mathcal{X}$. Due to the replacement (bootstrapped samples), objects can be duplicated in several subsamples. This procedure is known as \textit{bagging}. Each tree can use at most a number of features (\texttt{$N_{\rm F}$}), previously defined, randomly chosen for each tree to build themselves. This random subspace method guarantees that the classification of the algorithm is uniformly distributed over all trees and the entire forest works as a reliable classifier, instead of taking the classification from just one tree. Each bootstrapped subsample $\mathcal{S}_{\rm i}$ is used to build and train its respective $i$-th tree, passing through the tree as it grows. 

During the tree growth, at the $j$-th tree node, the Gini impurity is used to measure the goodness of the classification. It is defined as, 
\begin{equation}
\label{eq.Gini}
\mathcal{G}_j = 1 - \sum_{k=1}^{K} p_k^2,
\end{equation}
where $K$ is the total number of classes and $p_{\rm k}$ represents the probability of observing the $k$-th class which is estimated as the fraction of objects of that class at the node. After the split at the $j$-th node (considering a certain feature and threshold value), those elements go down to one of the two child nodes, called left ($l$) and right ($r$) nodes, building up their child population $N_{\rm l}$ and $N_{\rm r}$, respectively. The Gini impurity is then calculated at the child nodes assuming Eq. \ref{eq.Gini} for their respective object sets, i.e., $\mathcal{G}_{\rm l}$ and $\mathcal{G}_{\rm r}$. The Information Gain ($\mathcal{IG}$) is then obtained after the node split by the difference between impurities at the parent and child nodes, weighted by their respective number of objects, i.e.,
\begin{equation}
\label{eq.IG}
\mathcal{IG}_j = \mathcal{G}_j - f_l \mathcal{G}_l - f_r \mathcal{G}_r 
\end{equation}
where $f_{\rm l}$ and $f_{\rm r}$ represent the fractions of objects at the child node related to the parent node. Note that if the child nodes are pure, i.e., if there is only one class at each child nodes, the child Gini impurity values are zero and $\mathcal{IG}_j$ = 0 and it is not necessary to create new child nodes at that branch. These steps are taken to select the set of feature and threshold values that maximize $\mathcal{IG}$ and provides the best and most efficient combination of feature and threshold value for the split at the node. This procedure is repeated for child nodes which are not pure and new child node are created, making the tree grows as the purity increases. 

The tree growth happens until all leaves of the tree are pure or the minimum number of objects is reached, which in our case corresponds to two elements. Once all trees are built and trained using their respective subsamples $\mathcal{S}_i$, the RF is able to classify objects which are out of the training sample. Each tree outputs a classification label $\mathcal{Y}_{\rm i}$ for a certain object and the final class is denoted by the majority of the votes from all trees. In addition, the class probabilities are defined as the fraction of votes for each class. 

\begin{figure}
\includegraphics[scale=0.28]{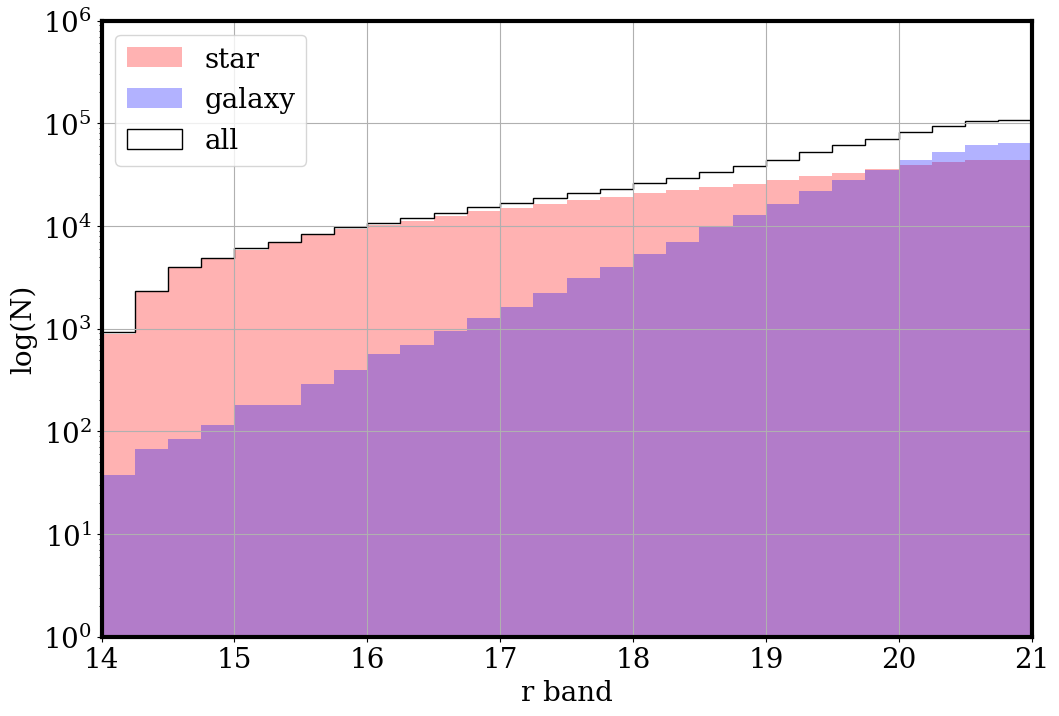}
\includegraphics[scale=0.28]{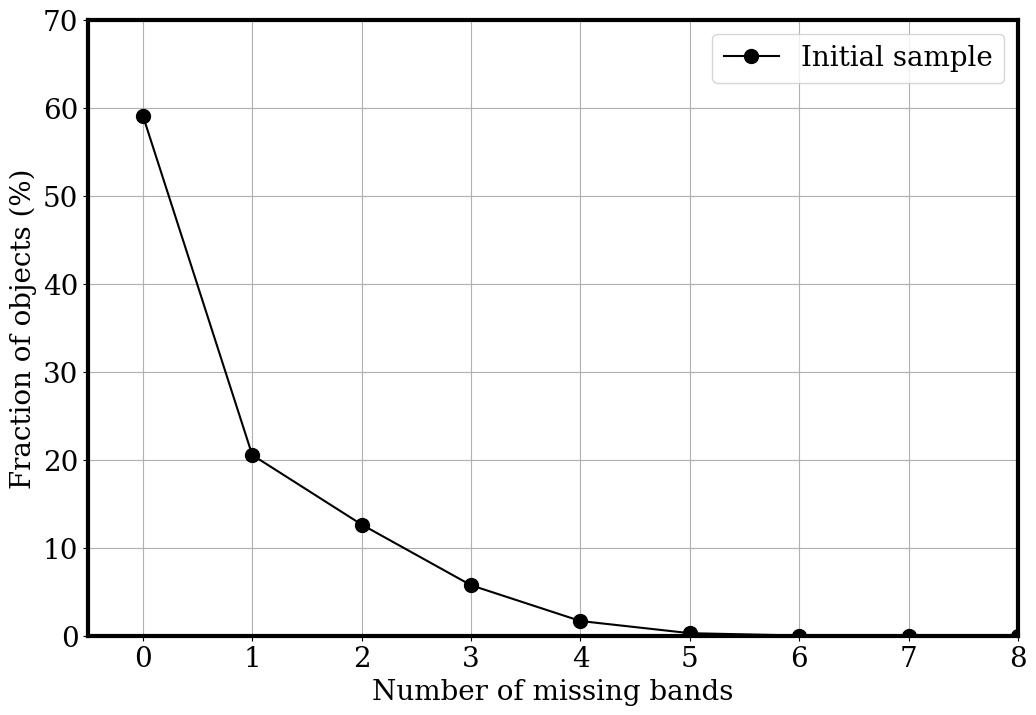}
\caption{Content and photometric coverage of the initial sample. Upper panel: the $r$ band distributions of objects classified as stars and galaxies according to the SDSS of the initial sample. Lower panel: the fraction of objects as a function of the number of missing bands. Both panels characterize our initial sample which was subsequently randomly split into training and test samples.}
\label{fig.train_test_samples}
\end{figure}

The RF algorithm has a few free parameters to be defined for the training process and classification. The parameter $n_{\rm tree}$ has the major role in the RF multi-dimensional classification. The performance of the algorithm was conservatively evaluated as function of the $n_{\rm tree}$ and it has been noticed that for $n_{\rm tree}$ $>$ 100, our results are statistically stable, i.e., presenting similar performances. Thus, we adopt the $n_{\rm tree}$ of 300 ($n_{\rm tree}$ = 300). Other parametrizations were defined as default, for instance, the number of features used by each tree is defined by \texttt{$\sqrt{N_{\rm F}}$} as well as the metric for the node split is the Gini impurity (\texttt{criterion=`gini'})\citep[][for more details]{Breiman2001}. All other scikit-learn parameters for the RF Classifier are set as default. 

The RF as well as other ML algorithms are quite sensitive to incomplete datasets. This issue, known as the missing feature problem, has to be addressed in S-PLUS since sometimes sources are not detected in all bands simultaneously (see Section \ref{data}) and therefore, there is not available/missing information. Thus, several approaches have been proposed to circumvent the missing data problem adopting, for instance, simple median values or imputation-procedures \citep{SchaferGraham2004,Reigeretal2010}. 

In this work, we have decided to be more conservative and train the RF algorithm for all combinations of missing features. In this way, our algorithm allows us to extract all information available for each object, independently of interpolations or assumptions which may not be correct. Initially, we consider the broad bands $g$, $r$, $i$ and $z$ bands as mandatory for the classification and all the others (uJAVA and narrow bands) may be missed. From 1 to 8 missing bands, the number of missing bands combinations are 255. Then 255 subsamples from the initial training sample were extracted to train 255 RFs which represent all missing bands combinations. It is important to mention that the missing bands problem depends mainly on the apparent magnitude, i.e., fainter objects systematically present higher number of missing bands. Moreover, narrow bands are more affected at the faint edge of the sample due to their lower photometric signal-to-noise when compared to the broad ones. Subsamples used to train cases of missed narrow bands are consequently shallower than the initial training sample. These 255 subsamples cover distinct regions in the magnitude/colour space and their contents span between 214k and the initial 362k, according to their number of missing bands. The training subsamples are systematically larger as the number of mandatory bands decreases. All objects in our training sample present mandatory measured $griz$ magnitudes. The training sample for this case is represented by the full case (362k). We trained RFs considering all 255 subsamples. The influence of the morphological parameters was evaluated by training RFs including (or not) those parameters. Therefore, we ended up with a total of 510 (255 missing bands combinations $\times$ 2 configurations, with or without morphology features) trained RFs for all cases of missing bands and morphology parameters inclusion. 

\section{RESULTS}
\label{results}

\subsection{Completeness, Purity and Misclassification}

After the training process, the performance of the trained RFs was evaluated using the test sample. We then define completeness as the fraction of correctly classified stars and purity is the ratio between the correctly classified stars and the total number of stars. Misclassification is the fraction of uncorrectly classified objects. Similarly, these parameters can also be implemented for galaxies. Since the initial training and test samples were randomly extracted from the S-PLUS database, it is expected that they occupy the same \textit{locus} in the colour-colour space and suffer from the missing bands problem similarly. 



\begin{figure*}
\includegraphics[scale=0.3]{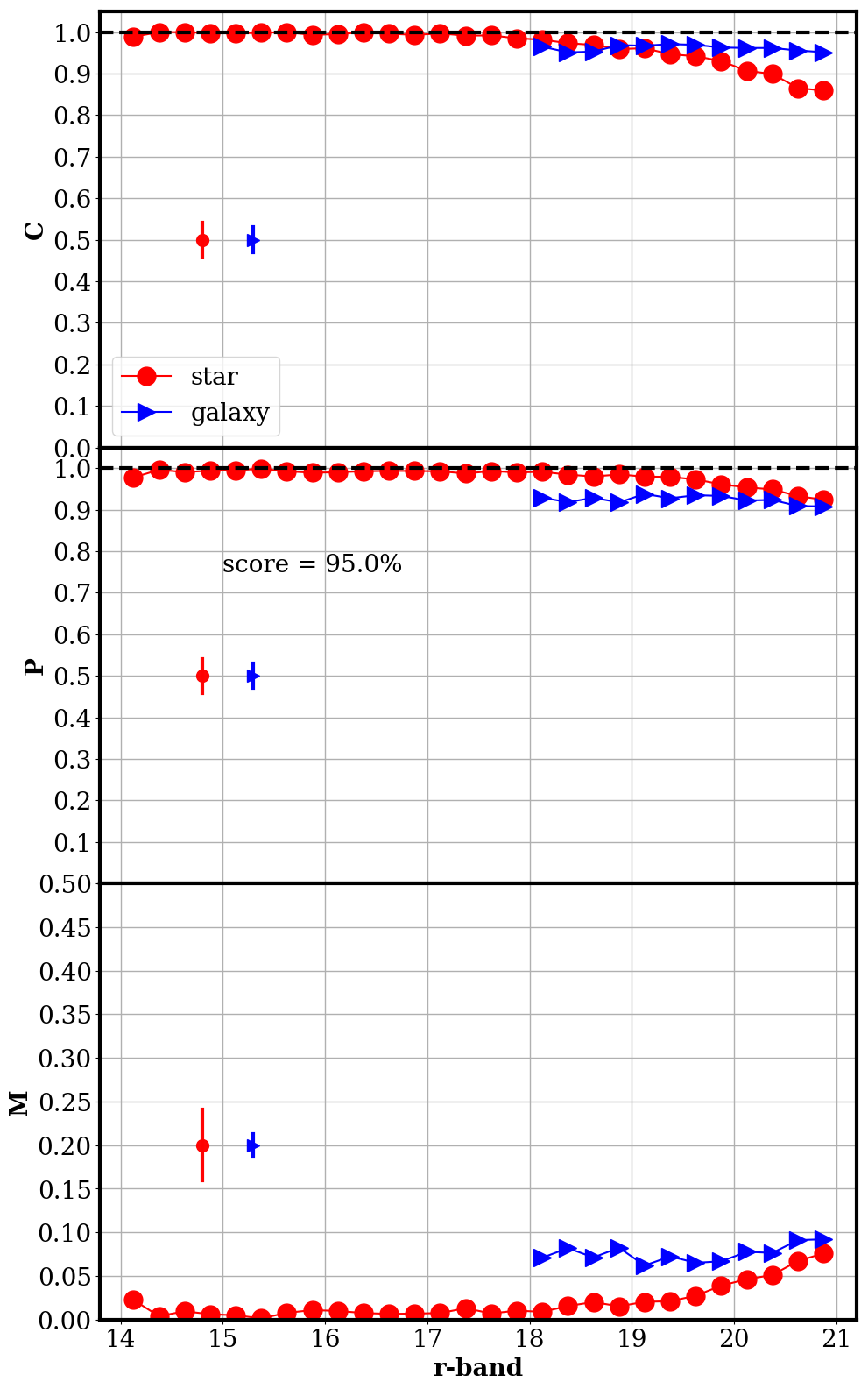}
\includegraphics[scale=0.3]{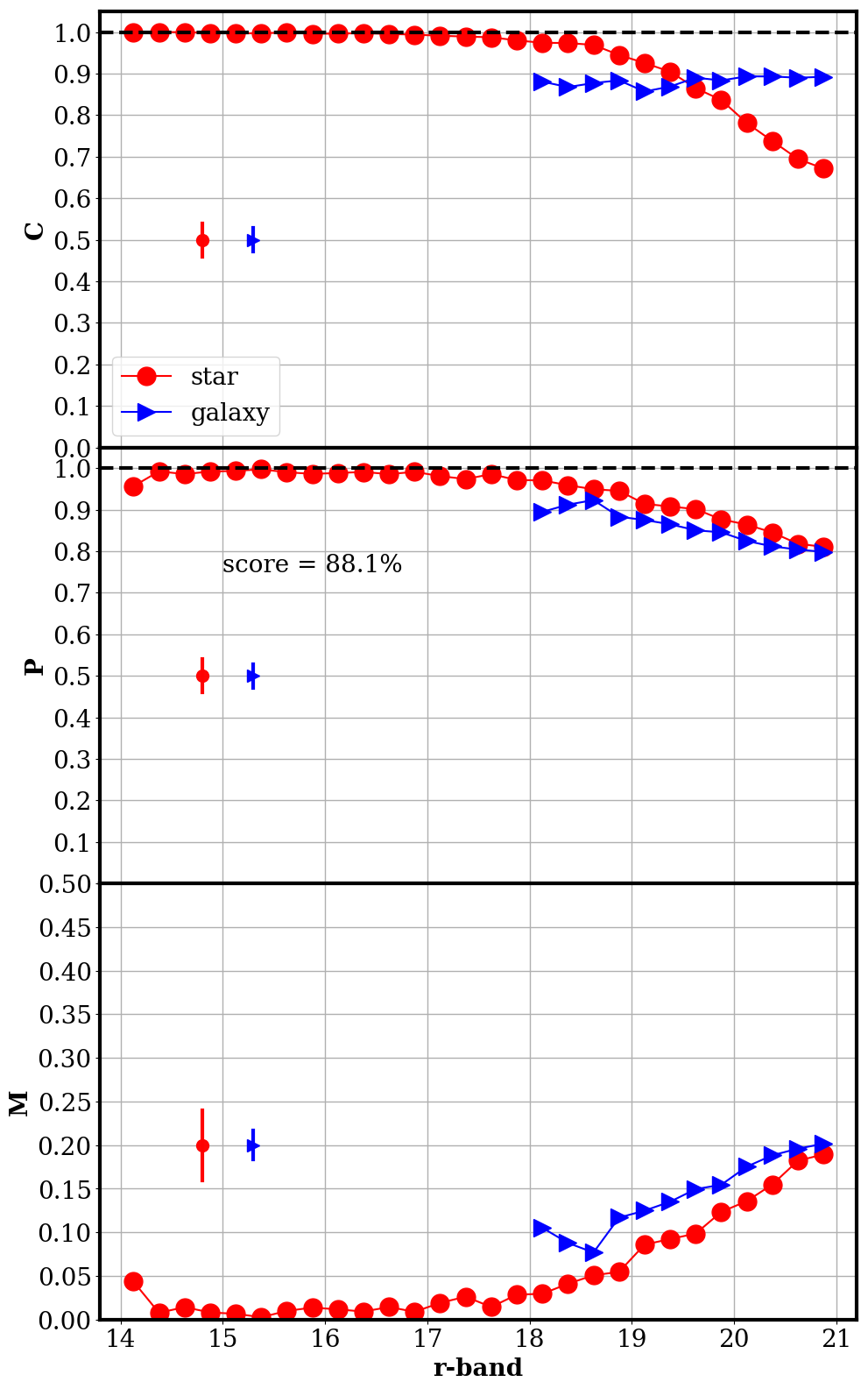}
\caption{The Completeness (upper panels), Purity (middle panels) and Misclassification (lower panels) as function of r-band for stars (circles) and galaxies (triangles) from the test sample (see text). The left and right figures represent the cases in which morphology is or is not included, respectively. The typical error bars are shown at the left of each panel.}
\label{fig.CPM}
\end{figure*}

Figure \ref{fig.CPM} shows the RF performance for the test sample, indicating the completeness, purity and misclassification of stars and galaxies as a function of the apparent r-band magnitude. Since S-PLUS/DR1 dataset up to now consists of a relatively small area in the sky, the number of bright galaxies is not enough to properly train the RF algorithm at that magnitude range. We decided to select solely those detections with a magnitude $r>$18 (see also Figure \ref{fig.train_test_samples}), leaving the classification of very bright galaxy for future datasets. 

When morphology is included in the features set (left panels in figure \ref{fig.CPM}), the completeness and purity reach 85\% or more for both classes down to $r$=21. The star completeness and purity systematically decline as they go to fainter magnitudes, as expected. However, the galaxy completeness and purity presented a roughly constant or at most slightly decreasing behaviour for fainter magnitudes. In average, the misclassification values for stars and galaxies are below 10\% for the majority of the magnitude range. On the other hand, the RF performance for features set composed only by S-PLUS magnitudes (right panels in Figure \ref{fig.CPM}) is systematically worse when compared to the morphology inclusion case. The star and galaxy completeness decreases down to 67\% and 89\% at the faint end, respectively. The misclassification presents values of 4\% for stars and 11\% for galaxies at $r$=18 and reaches values of 20\% at $r$=21 for both classes when the morphology is not included. 

The improved performance due to the inclusion of morphological parameters can be explained by the fact that galaxies frequently present {\rm FWHM/PSF} values larger than unity while stars are point-like sources ({\rm FWHM/PSF}$\sim$1), which represents a good discriminant to separate these two classes. Particularly, in the faint end of the sample, the morphological parameters are crucial to separate galaxies and stars. Moreover, the SED characterization becomes less informative as it goes to fainter magnitudes due to the photometric noise and the systematically increasing number of missing bands. The RF algorithm employs the morphology of objects to improve the classification particularly for fainter objects. For the worst case scenario, having only the broad bands $g$, $r$, $i$ and $z$, the accuracy is 94.8\% and 82.5\% when morphology is and not included, respectively. 

\subsection{Importance of Features}

The RF algorithm also allows to estimate the importance of features, i.e., how discriminant or decisive a certain feature is on the RF classification. To evaluate the relevance of each parameter, we define here the mean decrease accuracy. It is represented by the relative decrease in the initial accuracy ($acc_{\rm initial}$) when all trends related to a certain feature vanish. The feature trends are erased by shuffling the values related to the feature in the test sample. All objects are then re-classified after the feature shuffle and a new accuracy of performance is calculated ($acc_{\rm shuffled}$). Since there is no correlation with other features, the new accuracy is consequently lower. The importance of the feature is then represented by the relative decrease in accuracy, i.e., 
\begin{equation}
\label{eq.importance}
I = (acc_{\rm initial} - acc_{\rm shuffled}) / acc_{\rm initial}.
\end{equation}
Positive values of $I$ indicate that the feature has some importance for the classification. Null values indicate that the feature has no contribution in the algorithm classification. Note that the sum of all importance values ($I$) is not necessarily equal to unity. We have evaluated the importance of features for three different classification cases: i) only mandatory bands ($griz$), ii) 12 S-PLUS bands and iii) 12 S-PLUS bands plus including morphological parameters.

\begin{figure*}
\includegraphics[scale=0.4]{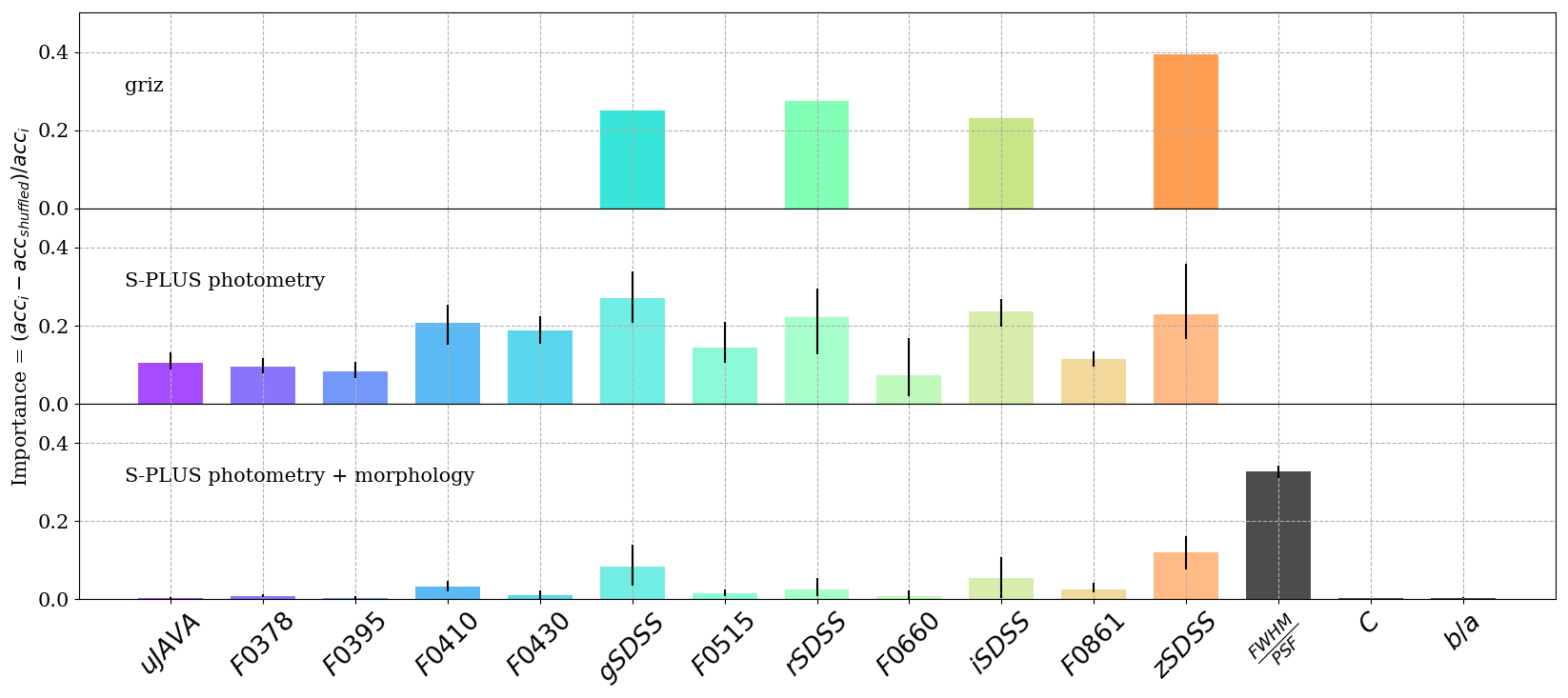}
\caption{The importance of features is calculated using Eq. \ref{eq.importance} for three set of features: only SDSS photometry (upper panel), S-PLUS photometry (middle panel) and S-PLUS photometry + morphology parameters (lower panel). The vertical black lines on the top of each bar represent 5\% and 95\% percentiles over all combinations of missing bands that include the feature.}
\label{fig.RF_importance}
\end{figure*}

Figure \ref{fig.RF_importance} shows the relative importance of features for the cases, i), ii) and iii). Regarding case i), we can notice the importance above 20\% for all bands. The $z$ band presented a slightly higher importance compared to the others. In case ii), we still notice the importance of mandatory bands in the classification. The $F0410$ and $F0430$ bands are strategically located at the redder region of the spectral feature $D_n4000$ at the rest-frame. Since the SED of stars are namely at the rest-frame and galaxies are always redshifted, these bands are important tracers of $D_n4000$ of stars while most galaxies do not have their $D_n4000$ detected in those bands due to their redshifts. As a consequence, its importance is slightly higher when compared to the other narrow bands. Case iii) shows the inclusion of morphological parameters in the classification. Our result shows that the {\rm FWHM/PSF} presents by far the largest importance, being more than 30\%. It indicates the relevance of the morphology, particularly how extended objects appear to be classified as stars and galaxies. Besides this result, we can still see minor order importance for $z$ and $g$. Previous works in the literature already pointed out to the importance of {\rm FWHM} in classifying objects into stars and galaxies \citep[e.g., ][]{LopezSanJuanetal2018}.  

\subsection{The RF performance in colour-colour space}

\begin{figure*}
\includegraphics[scale=0.22]{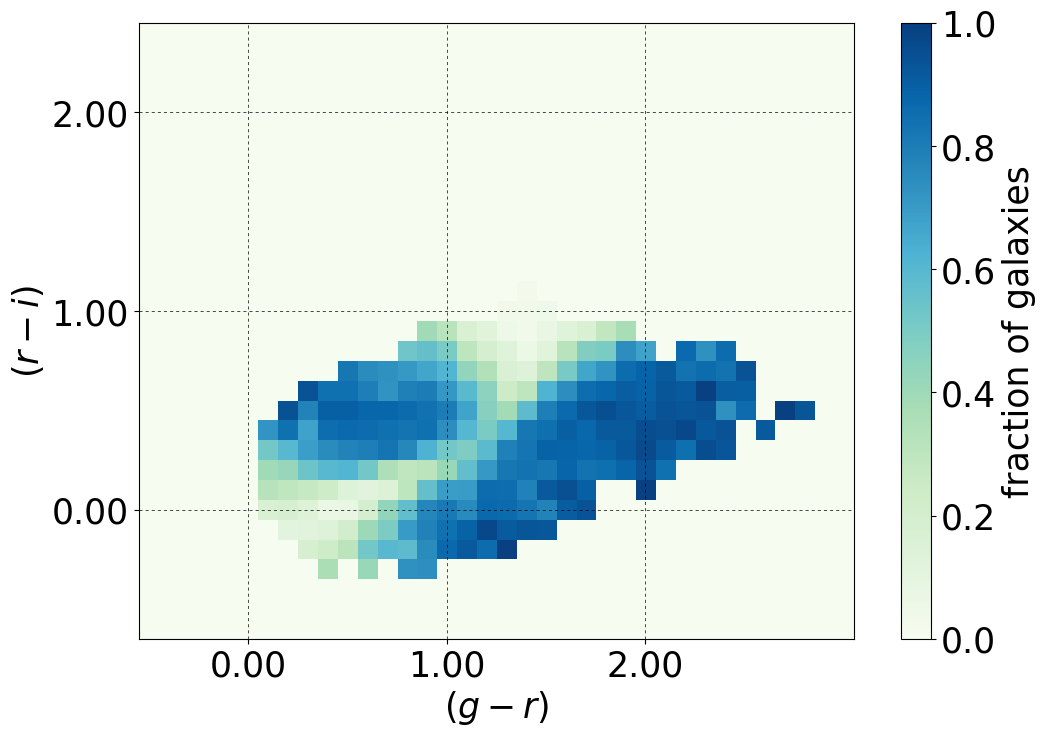}
\includegraphics[scale=0.22]{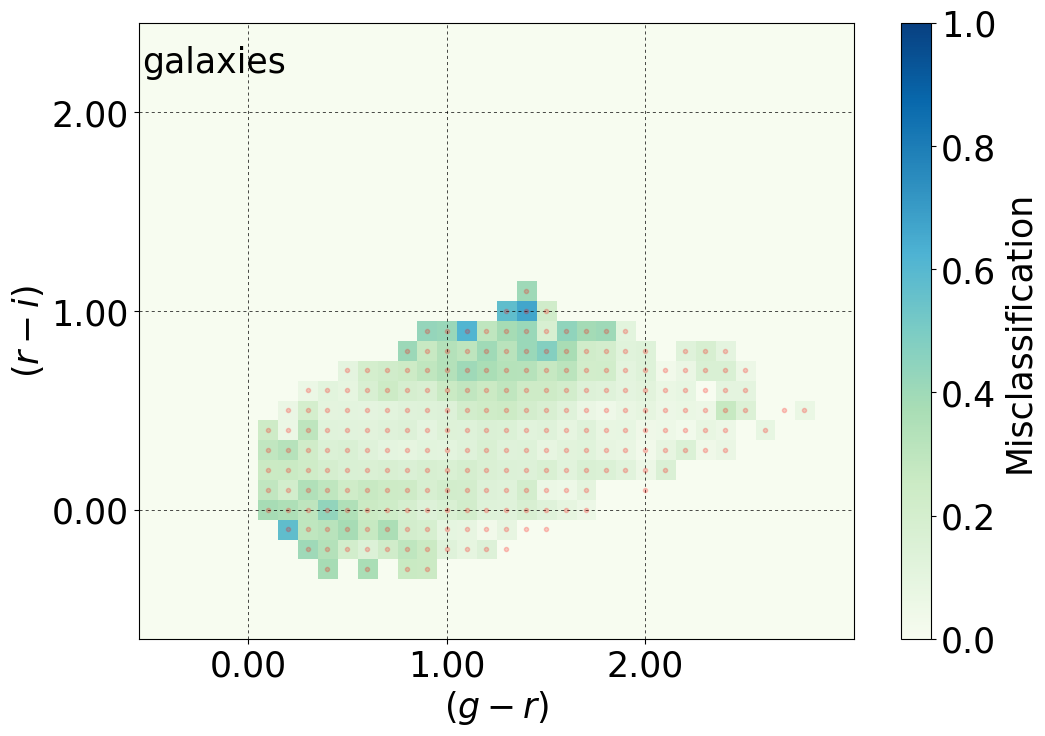}
\includegraphics[scale=0.22]{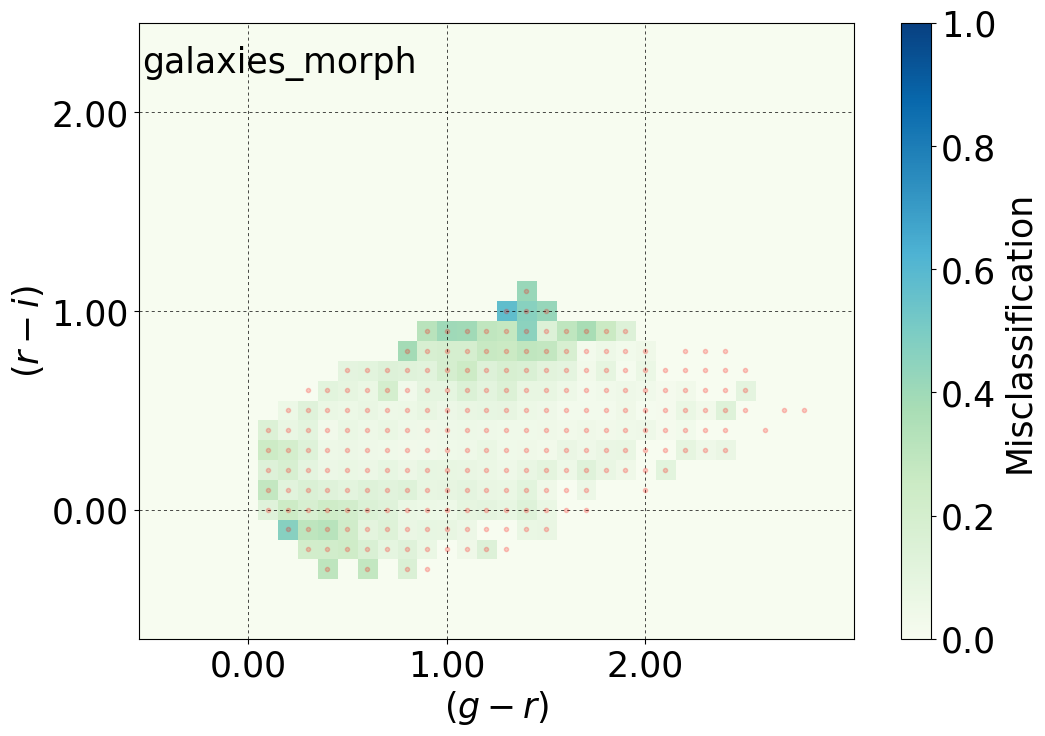}

\includegraphics[scale=0.22]{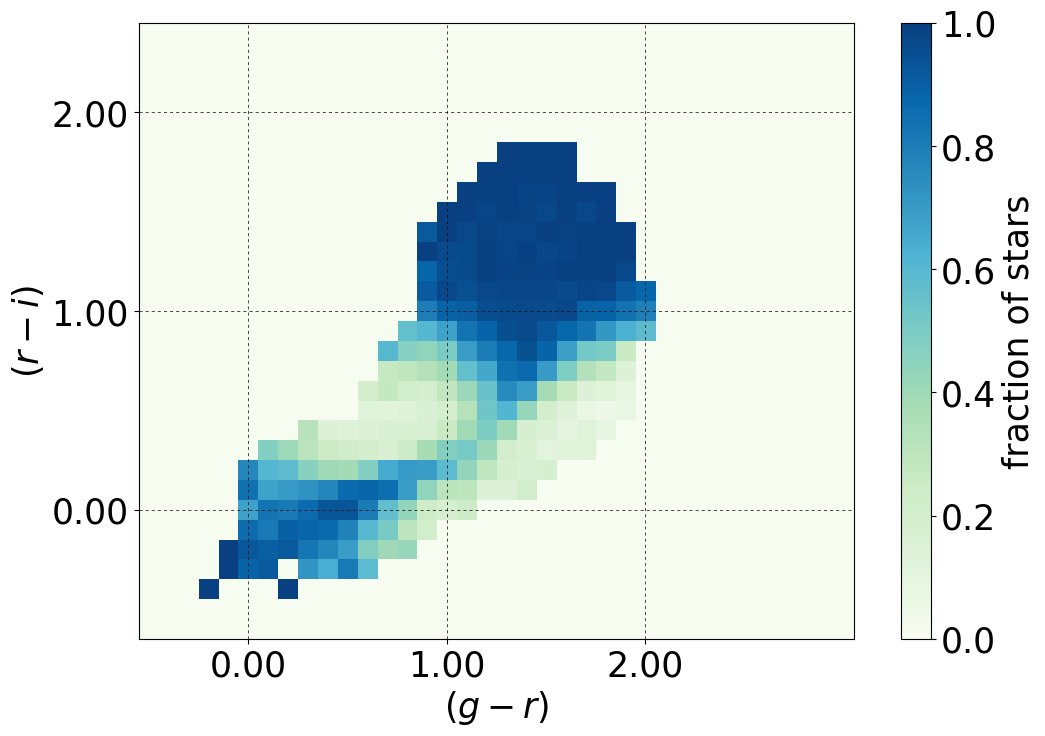}
\includegraphics[scale=0.22]{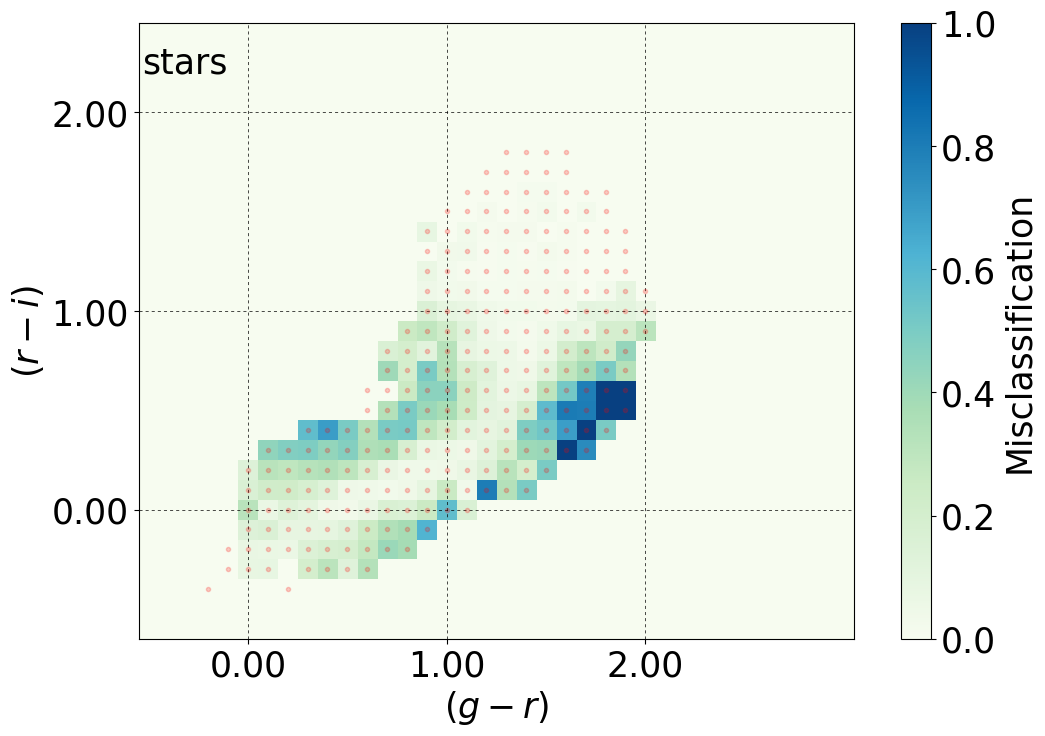}
\includegraphics[scale=0.22]{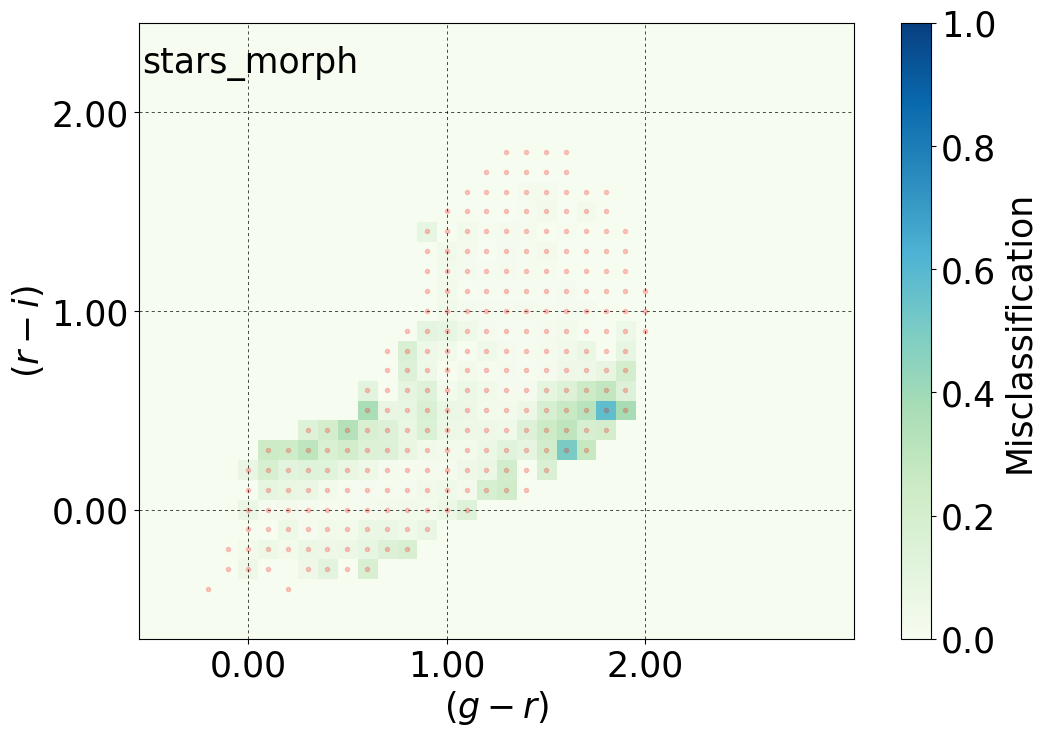}

\caption{The colour-colour diagram $(g-r)$ versus $(r-i)$ for galaxies (upper panels) and stars (lower panels), showing the fraction of objects classified as galaxies and stars (left), misclassification using only S-PLUS photometry (middle) and the misclassification using S-PLUS photometry plus morphology (right). The misclassification was evaluated only for cells in the diagram which correspond to at least ten objects (red points in the middle and right panels).}
\label{fig.colour_colour_M}
\end{figure*}

The performance of the RF algorithm in the colour-colour space was also evaluated in this work, presenting regions with higher or lower performance as function of two broad band colours. Mandatory bands are chosen here because are not significantly biased due to the photometric depth. Figure \ref{fig.colour_colour_M} shows the colour-colour space: $(g-r)$ versus $(r-i)$ containing galaxies and stars samples. The fraction of galaxies (upper panels), stars (lower panels) and their misclassification by using (or not) the morphology parameters are indicated by the colour maps. First, one can notice that stars and galaxies overlap in the colour-colour space. However, they prevail in distinct regions in such diagram (left panels). By comparing the fraction of galaxies or stars with their misclassification map in the colour-colour space, we notice a correlation between low fraction of objects and high misclassification. In regions which present galaxy fractions of 20\% or lower, the misclassification of galaxies is higher than other regions. This correlation can be explained the low statistical significance of number of galaxies. In other words, when there is a large number of stars and low number of galaxies in a specific region in the diagram, the misclassification of galaxies at that region is more noisy due to the low number of objects of that class. The misclassification at those regions are shown in Figure \ref{fig.colour_colour_M}, including only S-PLUS photometry (middle panels). Including morphological information in the RF algorithm, particularly the parameter {\rm FWHM/PSF}, the misclassification decreases as a whole. In particular, regions with relatively high misclassification values due to the low number of objects of a certain class have their performance (lower misclassification) improved when the morphology is included (right panels).

Notably, galaxies with $(r-i)\sim 1$ and $(g-r)\sim 1.5$ do not present a significant improvement on classification with the morphology inclusion. At this region, one can notice that the number of galaxies is quite low when compared to the other ones, being probably the reason of no improvement. 

\subsection{What about QSOs?}

Since QSOs are luminous, extragalactic and point-source objects, they are commonly classified as stars in most of star/galaxy classification algorithms. We evaluate our classification for these objects using a matched sample between the S-PLUS and the SDSS spectroscopic QSO confirmed objects, containing 8571 objects. Note that this spectroscopic sample is not complete due to the SDSS spectroscopic completeness when compared to S-PLUS photometric depth. However, this exercise is useful to show how our algorithm classifies QSOs.

The classification including morphology indicates that 18.9\% of QSOs are classified as galaxies and 81.1\% as stars. If we do not include morphological parameters, these numbers are 23.7\% and 76.3\%, respectively. The classification of QSOs as extragalactic objects is slightly better for the only photometry case but the majority of the QSOs are still classified as stars. It is important to mention that our classification reproduces the SDSS one. As the SDSS classifies most of QSOs as stars due to their point-source morphology, this caveat was expected. We present here a caveat of the algorithm on the QSO classification, reflecting a limitation of the training sample. On the other hand, it is not the scope of this work to classify QSOs. This problem will be tackled in other S-PLUS articles in the future (Nakazono et al., Queiroz et al., in preparation).

\section{Applications}
\label{sec.application}
\subsection{The red galaxy sample contaminated by stars}


Our results can also provide an estimate of the expected contamination rate in red galaxy samples by stars as a function of the redshift and colour. The assessment of stellar contamination is very important for cosmological studies such those on galaxy clustering. Such studies may employ contamination corrections on galaxy samples by specific types of object to remove systematic effects \citep[e.g.][]{Xavieretal2018,Peacock18mn}. Moreover, the contamination estimate may help devise redshift and colour cuts that reduce the impact of such effects.

\begin{figure}

\includegraphics[scale=0.32]
{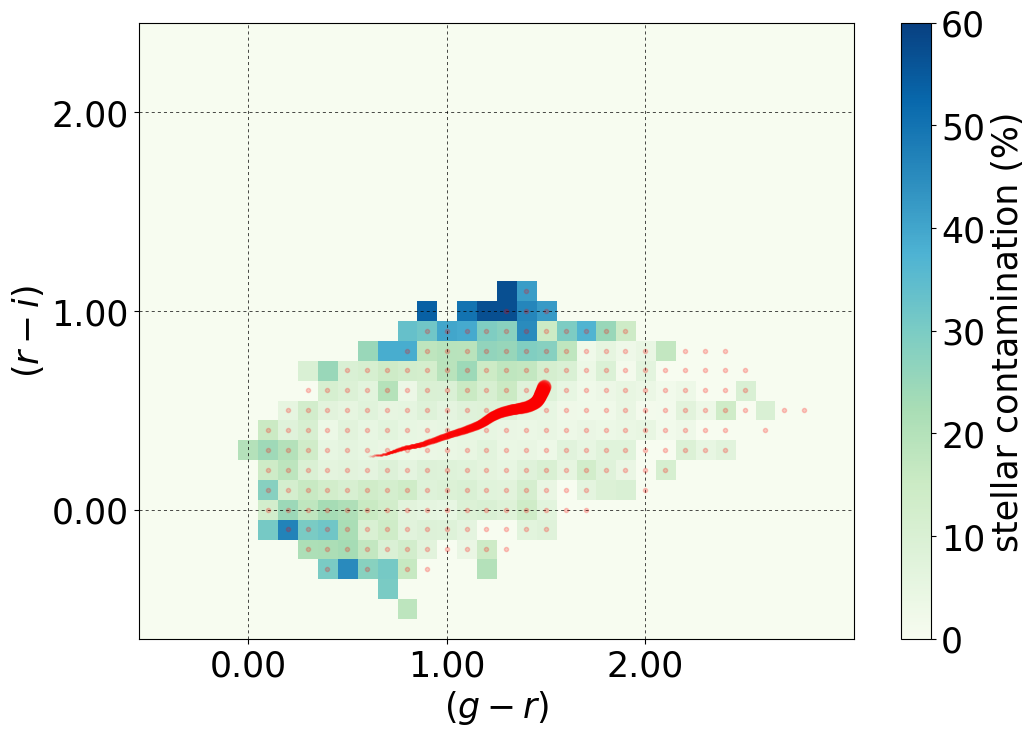}
\includegraphics[scale=0.34]{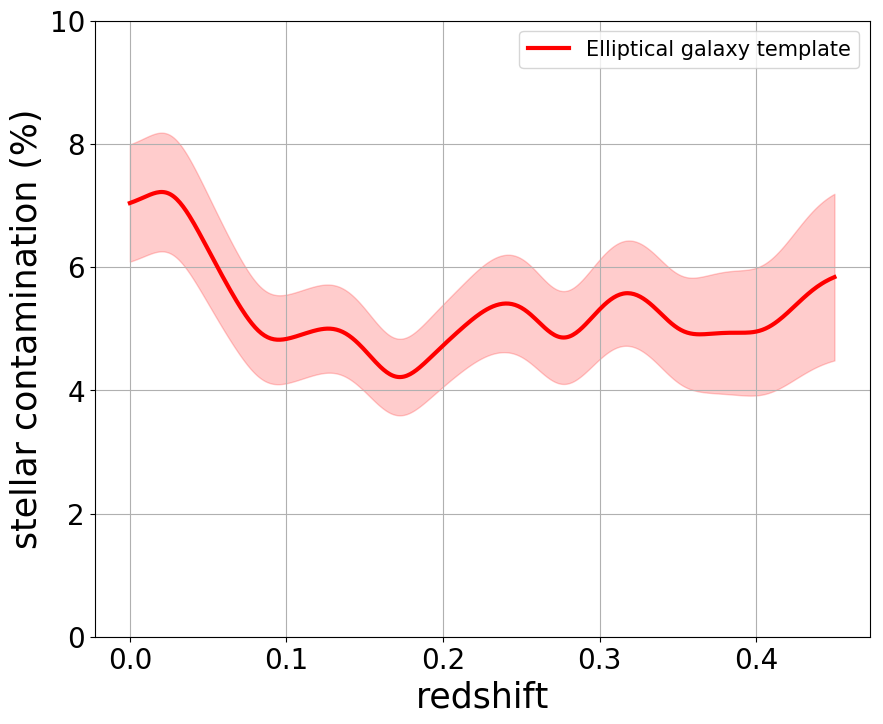} 
\caption{Upper panel: the stellar contamination in the colour-colour space $(g-r)$ versus $(r-i)$ associated to galaxies. It shows the redshift track of a red galaxy template convolved with the broad bands $g$, $r$ and $i$ up to redshift $z=0.45$. Lower panel: the expected contamination by stars for a red galaxy sample as a function of the redshift. The shaded area represents 1$\sigma$ uncertainties.}
\label{fig.red_galaxy_contamination}
\end{figure}

We have estimated the stellar contamination fraction $f_{\mathrm{star}}$ in a galaxy sample using the following equation:

\begin{equation}
f_{\mathrm{star}} = \frac{N^{\mathrm{g}}_{\mathrm{fake}}}{N^{\mathrm{g}}_{\mathrm{fake}}+N^{\mathrm{g}}_{\mathrm{real}}},
\end{equation}
where $N^{\mathrm{g}}_{\mathrm{fake}}$ is the number of stars classified as galaxies and $N^{\mathrm{g}}_{\mathrm{real}}$ is the number of correctly classiflied galaxies. Our estimate is presented as a function of colour in Figure \ref{fig.red_galaxy_contamination} (upper panel). This estimate can serve as a guidance for the implementation of colour cuts aimed at improving the purity of a galaxy sample. 

As an example, we have considered the case of luminous red galaxies (LRGs), which are frequently used in clustering analysis due to their high bias that eases the tracking of matter density contrasts. In order to simulate the expected colour evolution of an LRG in our colour-colour diagrams, we made use of the reddest galaxy template from the library of SED models of the Bayesian Photometric Redshift (BPZ2.0) code \citep{Benitez2000, Molinoetal2017, Molinoetal2018}. We created a grid of SED models by redshifting the reddest galaxy template from $z=0.00$ to $z=0.45$ at steps of $dz=0.01$. Then each SED model was convolved with our filter system in order to derive the expected model magnitudes in S-PLUS. This red galaxy track in the colour-colour diagram is presented in Figure \ref{fig.red_galaxy_contamination}, showing that this type of galaxy populates a region of low contamination rate. 


Figure \ref{fig.red_galaxy_contamination} (lower panel) shows the interpolated contamination rate for red galaxies as a function of redshift. We can see that our classification method reaches 5\% contamination over the whole S-PLUS redshift range. As shown in \citet{Peacock18mn} and \citet{Xavieretal2018}, such contamination level is low enough to allow the extraction of reliable cosmological parameters from the data. 

\subsection{The S-PLUS extragalactic point-source catalogue}
\label{pointsource_extragal_cat}

The final product of this work is a star/galaxy classification for the S-PLUS catalog. As a subproduct, it provides a catalog of extragalactic point-source objects that are candidate to high-z galaxies, QSOs, compact HII galaxies and Ultra Compact Dwarf galaxies (UCDs). The classification only based on the S-PLUS photometry identifies objects classified as extragalactic ones, independently of the morphology (see Section \ref{sec.RF}). 

Some constraints were imposed to select point-source and extragalactic objects:

\begin{itemize}
\item $14<r<21$
\item $z_b$$>$0.002 
\item 1.0$<${\rm FWHM/PSF}$<$1.5
\item $P_{gal}$$>$0.5 (classified as galaxies)
\end{itemize}
where $P_{gal}$ represents the probability of a certain object to be classified as a galaxy using the RF algorithm only based on 12 S-PLUS photometric bands. The photometric redshifts ($z_b$) were derived from \cite{Molinoetal2019}. Our initial catalogue contains roughly $\sim$18k objects. This extragalactic point-source catalogue will be explored in the future by the S-PLUS collaboration.

\section{Conclusions}
\label{conclusions}

We present a star/galaxy classification for the S-PLUS project, based on the RF algorithm. This approach uses the 12 S-PLUS photometric bands as input for our classification. To evaluate the influence of the morphological information on the classification, we used two set of features. The first one consisting of only 12 all S-PLUS bands while the second one is represented by the 12 photometric bands and the morphological parameters, then we included and evaluated the impact of the following morphological parameters in the classification: full width at half maximum divided by the representative {\rm PSF} of the field ({\rm FWHM/PSF}), ellipticity ($b/a$) and concentration ($C$) at $r$ band. The matched sample between the S-PLUS/S82 and SDSS/S82 was split into training and test samples, presenting 724k and 200k objects, respectively. To circumvent the missing bands issue in the S-PLUS, we trained several RFs for all combinations of missing bands. In total, 510 RFs were trained for all missing bands combinations and morphological information included or not. Our conclusions can be summarized as follows:

\begin{itemize}
\item Morphological information, in particular {\rm FWHM/PSF}, is crucial for the star/galaxy classification. The performance of the algorithm with the inclusion of morphological information is considerably improved. The perfomance of the algorithm is improved from an accuracy of 88.1\% to 95.0\% with the inclusion of the morphological information.

\item The use of broad bands presented slightly larger importance when compared to the narrow ones due to their higher signal-to-noise, with one exception, the narrow band J0410 for the case when only 12 S-PLUS photometry is considered. This is one of the bands that traces the $D_n4000$ at the rest-frame, showing its importance to separate Galactic and extragalactic objects. Moreover, the {\rm FWHM/PSF} presented the highest importance with more than 33\% among all other features. 

\item We also presented the misclassification in the colour-colour diagram, showing the misclassification of stars and galaxies as function of two broad band colours, $(g-r)$ versus $(r-i)$. The classification improves due to the parameter {\rm FWHM/PSF}, notably at the \textit{locii} where the misclassification is relatively high. Regions where the classification is relatively uncertain when using only 12 S-PLUS bands had their misclassification decreased by including morphology information. One of the outputs is the stellar contamination (or misclassification) of a red galaxy sample as a function of redshift. This result can be useful for extragalactic studies. 

The final product of this work is the classification of objects in the S-PLUS project as stars or galaxies. Also, we emphasize the importance of stellar contamination for extragalactic studies... etc.

\item Another data product of this star/galaxy separation algorithm is the extragalactic point-source catalogue. The classification using only 12 S-PLUS bands and additional constraints on photo-z and {\rm FWHM/PSF} values generates a sample with $\sim$18k objects, candidates for extragalactic point-source objects.    

\end{itemize}

\section{Acknowledgements}
 
MVCD thanks the FAPESP scholarship process number 2014/18632-6 and 2016/05254-9. LS acknowledges the financial support of the FAPESP scholarship process number 2016/21664-2. We also thank the FAPESP thematic project number 2012/00800-4.
FRH thanks FAPESP scholarship number 2018/21661-9. ACS acknowledges funding from the brazilian agencies \textit {Conselho Nacional de Desenvolvimento Cient\'ifico e Tecnol\'ogico} (CNPq) and the Rio Grande do Sul Research Foundation (FAPERGS) through grants CNPq-403580/2016-1, CNPq-310845/2015-7, PqG/FAPERGS-17/2551-0001. AAC acknowledges support from FAPERJ (grant E26/203.186/2016) and CNPq (grants 304971/2016-2 and 401669/2016-5). JLNC is grateful for financial support received from the GRANT PROGRAM  FA9550-18- 1-0018 of the Southern Office of Aerospace Research and development (SOARD), a branch of the Air Force Office of the Scientific Research International Office of the United States (AFOSR/IO).

This work has made use of the computing facilities of the Laboratory of Astroinformatics (IAG/USP, NAT/Unicsul), whose purchase was made possible by the Brazilian agency FAPESP (grant 2009/54006-4) and the INCT-A.

\nocite{*}
\bibliographystyle{mn2e}

\label{lastpage}
\end{document}